\documentclass[12pt]{article}
\usepackage{graphicx}
\def \b{{\cal B}}
\def \bea{\begin{eqnarray}}
\def \beq{\begin{equation}}

\def \eea{\end{eqnarray}}
\def \eeq{\end{equation}}

\begin{document}
\Large
\centerline{\bf Low-Mass Baryon-Antibaryon Enhancements}
\centerline{\bf in $B$ Decays
\footnote{Enrico Fermi Institute preprint EFI 03-11, hep-ph/0303079.
To be submitted to Phys.\ Rev.\ D.}}
\normalsize
\bigskip
 
\centerline{Jonathan L. Rosner~\footnote{rosner@hep.uchicago.edu}}
\centerline {\it Enrico Fermi Institute and Department of Physics}
\centerline{\it University of Chicago, 5640 S. Ellis Avenue, Chicago, IL 60637}
\bigskip
 
\begin{quote}

The nature of low-mass baryon-antibaryon enhancements seen in $B$ decays is
explored.  Three possibilities include (i) states near threshold as found in
a model by Nambu and Jona-Lasinio, (ii) isoscalar states with $J^{PC} =
0^{\pm +}$ coupled to a pair of gluons, and (iii) low-mass enhancements favored
by the fragmentation process.  Ways of distinguishing these mechanisms using
angular distributions and flavor symmetry are proposed.

\end{quote}

\noindent
PACS Categories: 13.25.Hw, 14.40.Nd, 13.75.Cs, 14.40.Cs

\bigskip

\centerline{\bf I.  INTRODUCTION}
\bigskip

In many decays $B \to \bar D^{(*)} N \bar N$ \cite{CLDNN,BeDNN}, $B^+ \to K^+ p
\bar p$ \cite{Kpp}, and $B^0 \to \bar \Lambda p \pi^-$ \cite{Lppi}, the
baryon-antibaryon effective mass peaks at very low values.  Even more
pronounced peaking at low baryon-antibaryon effective mass has now been
observed in the radiative decay $J/\psi \to \gamma p \bar p$ \cite{gpp}.
The effective mass distribution in this last process is so sharply peaked near
threshold that for an S-wave the data can been interpreted in terms of a
$p \bar p$ bound state with $M = 1859^{+3+5}_{-10-25}$ MeV/$c^2$ and $\Gamma <
30$ MeV (90\% c.l.).  For a P-wave a fit \cite{gpp} yields a state just at $p
\bar p$ threshold:  $M = 1876.4 \pm 0.9$ MeV/$c^2$, $\Gamma = 4.6 \pm 1.8$ MeV.
There are numerous earlier claims for such states (see, e.g., \cite{Bridges86}
and \cite{Dalkarov97}), but not much unanimity on their properties.  An
enhancement near threshold is seen in $p \bar p \to e^+ e^-$ \cite{Bar94},
while various multi-particle production processes such as $e^+ e^- \to $
hadrons \cite{Ant98}, $e^+ e^- \to 6 \pi$ \cite{SLO}, and diffractive
photoproduction of $6 \pi$ \cite{Frab01} show dips at $p \bar p$ threshold.

Theoretical investigations of baryon-antibaryon bound states date back to the
proposal of Fermi and Yang \cite{FY} to make the pion out of a
nucleon-antinucleon pair.  The model of Nambu and Jona-Lasinio \cite{NJL},
which is constructed to give a nearly zero-mass pion as a fermion-antifermion
bound state, also has a scalar resonance of twice the fermion mass.
Enhancements in the baryon-antibaryon channel near threshold are expected
on the basis of duality arguments \cite{JRbb,HHdd,JRdd} and by comparison
with the systematics of resonance formation in meson-meson and meson-baryon
channels \cite{JRcrf}.  A historical survey of bound states or resonances
coupled to the nucleon-antinucleon channel is given in Ref.\ \cite{Richard}.
Gluonic states can couple to baryon-antibaryon
channels of appropriate spin and parity.  Recent discussions of $B$ decays
involving baryon-antibaryon pairs include Refs.\ \cite{ID98,HS,CHT,CY,CH}.

In the present note we suggest some tests that may be useful in sorting out the
various interpretations of the observed effects near or below baryon-antibaryon
thresold.  Gluonic states with $J^{PC} = 0^{-+}$ can couple to isoscalar
$p \bar p$ pairs in a $^1S_0$ state, while those with $J^{PC} = 0^{++}$ can
couple to such pairs in a $^3P_0$ state.  The decays of both such states
into $p \bar p$ are isostropic.  Fragmentation-based effects need not (and,
we shall argue, should not) lead to such isotropy.  The decays of gluonic
states should be flavor-symmetric, while fragmentation products need not be.
The decay $B^\pm \to p \bar p K^\pm$ may occur through a similar mechanism
which gives rise to $B^\pm \to \eta' K^\pm$ and $B^0 \to \eta' K^0$, involving
the emission of two gluons by a penguin diagram.

We discuss gluonic mechanisms in Section II and fragmentation mechanisms in
Section III.  Section IV contains some more general remarks about the
possibility of observing baryon-antibaryon and other exotic resonances in $B$
decays, while Section V concludes.
\bigskip

\centerline{\bf II.  GLUONIC MECHANISMS}
\bigskip

Decays of the form $B \to K + X$ receive an important contribution from
a ``flavor-singlet penguin'' amplitude.  Here the fundamental subprocess
is $\bar b \to \bar s + g + g$, where $g + g$ stands for a pair of gluons
or a non-perturbative structure with vacuum quantum numbers.  The need for
this amplitude was anticipated \cite{GR96,DGR96} before it appeared
experimentally in the decays $B \to \eta' K$ \cite{CLEO98}.  The $\eta'$,
being largely a flavor-singlet meson, couples strongly to a pair of gluons
with $J^{PC} = 0^{-+}$.  A flavor-singlet penguin contribution which boosts
that of the ordinary penguin amplitude by as little as 50\% suffices to
explain the observed decay rate \cite{DGR97,CR01}.  Taking account of
interference with the ordinary penguin amplitude (whose importance is
considerable; see the arguments by Lipkin \cite{Lpen}), the branching ratio of
$B^+$ to $\eta' K^+$ due to the singlet penguin (sp) alone was estimated
to be \cite{CR01}
\beq \label{eqn:sp}
\b(B^+ \to \eta' K^+)|_{\rm sp} \ge 1.1 \times 10^{-5}~~~.
\eeq
The inequality becomes an equality if the singlet and ordinary penguin
interfere constructively.
We shall use this result to estimate the value of $\b(B^+ \to p \bar p K^+)$  
due to a gluonic mechanism.

The decays $B^+ \to p \bar p K^+$ and $J/\psi \to \gamma p \bar p$ both appear
to be dominated by a $p \bar p$ bound state whose mass we shall take to be
that for the $0^{-+}$ (S-wave) fit presented in Ref.\ \cite{gpp}, or
1859 MeV/$c^2$.  We shall denote this state by $E$.  We assume
\beq
\frac{\b(B^+ \to E K^+)|_{\rm sp}}{\b(B^+ \to \eta' K^+)|_{\rm sp}} =
\frac{\b(J/\psi \to \gamma E)}{\b(J/\psi \to \gamma \eta')}
\eeq
modulo phase space corrections.  Since $E$ is assumed to be spinless, the
$B^+ \to E K^+$ and $B^+ \to \eta' K^+$ decays are characterized by S-wave
kinematic factors proportional to the first power of the center-of-mass
(c.m.) momenta $p^*$, with $p^*_{B^+ \to E K^+} = 2282$ MeV/$c$ and $p^*_{B^+
\to \eta' K^+} = 2528$ MeV/$c$, respectively.  The magnetic dipole (M1)
$J/\psi$ decays contain kinematic factors proportional to $p^{*3}$, with
$p^*_{J/\psi \to \gamma E} = 990$ MeV/$c$ and $p^*_{J/\psi \to \gamma \eta'} =
1400$ MeV/$c$.  We use the branching ratios \cite{PDG} 
\beq \label{eqn:Jpsibrs}
\b(J/\psi \to \gamma p \bar p) = (3.8 \pm 1.0) \times 10^{-4}~~,~~~
\b(J/\psi \to \gamma \eta') = (4.31 \pm 0.30) \times 10^{-3}~~~,
\eeq
whose ratio is $0.088 \pm 0.024$, to calculate
\beq
\frac{\b(B^+ \to E K^+)|_{\rm sp}}{\b(B^+ \to \eta' K^+)|_{\rm sp}} =
\frac{2282}{2528} \left( \frac{1400}{990} \right)^3 (0.088 \pm 0.024)
= 0.23 \pm 0.06~,
\eeq
or, combining this result with Eq.\ (\ref{eqn:sp}),
\beq
\b(B^+ \to E K^+)|_{\rm sp} \ge (2.5 \pm 0.7) \times 10^{-6}~~~.
\eeq
This lower bound is to be compared with the observed branching ratio
\cite{Kpp}
\beq \label{eqn:Kpp}
\b(B^+ \to p \bar p K^+) = (4.3^{+1.1}_{-0.9} \pm 0.5) \times 10^{-6}~~~.
\eeq
Thus, the singlet penguin amplitude is expected to provide a fair fraction
of the observed final state.  Reasons for a shortfall could be:
(1) the singlet penguin amplitude is larger than its lower bound based on Eq.\
(\ref{eqn:sp}); (2) there could be some additional contribution from another
$p \bar p$ partial wave, such as $^3P_0$ ($J^{PC} = 0^{++}$); (3) there could
be a contribution from the fragmentation mechanism to be discussed in the
next Section.

The angular distribution of the photon in $J/\psi \to p \bar p \gamma$ is
found to be compatible with the $1 + \cos^2 \theta^*$ for expected if the
$p \bar p$ system is in a state with $J^{PC} = 0^{-+}$ \cite{gpp}.  Here
$\theta^*$ is measured with respect to the beam direction in the $e^+ e^-$
c.m.  The same angular distribution is expected for a $0^{++}$ ($^3P_0$)
$p \bar p$ state.  The two possibilities could be distinguished from one
another by measuring the photon polarization, e.g., in the Dalitz process
$J/\psi \to p \bar p e^+ e^-$.

If the $p \bar p$ system is in a $J=0$ final state (whether $0^{-+}$ or
$0^{++}$), the band for the low-mass $p \bar p$ enhancement in the Dalitz
plot for $J/\psi \to p \bar p \gamma$ should be uniformly populated.
A similar remark holds for the $p \bar p$ system in $B^+ \to p \bar p K^+$
if the singlet penguin mechanism is dominant.  In such a case one expects
the c.m.\ momentum distributions of $p$ and $\bar p$ to be identical.  As we
shall argue, this is not necessarily the case in a fragmentation picture.

When the $p \bar p$ system is produced through a pair of gluons (or any such
flavorless state), one should expect isospin symmetry to give the same
production rate for an $n \bar n$ system.  This prediction is difficult to
test.  In the limit of flavor-SU(3) symmetry one would also expect the same
rate for $B_8 \bar B_8$, where $B_8$ is any member of the baryon octet, but
SU(3)-breaking could alter this prediction considerably.  For example, the
proposed $p \bar p$ bound state at 1859 MeV is far below $\Lambda \bar
\Lambda$, $\Sigma \bar \Sigma$, or $\Xi \bar \Xi$ threshold, reducing the
likely branching ratios when $p \bar p$ is replaced by a hyperon-antihyperon
pair.  Some nucleon-antinucleon bound states proposed to exist near or below
threshold have $I=1$ \cite{Dalkarov97}, and could not be identified with the
gluonic effect we are proposing.
\bigskip

\centerline{\bf III.  FRAGMENTATION MECHANISMS}
\bigskip

The gluonic mechanism of the previous Section is unlikely for certain $B$
decays involving low-mass $p \bar p$ states.  A singlet penguin mechanism
cannot account for such decays as $B^0 \to \bar D^0 p \bar p$ and $B^0 \to
\bar \Lambda p \pi^-$.  Instead, a fragmentation picture is appealing; this
may also play a role in $B^+ \to K^+ p \bar p$.

Let us consider the example of $B^0 \to \bar D^0 p \bar p$.  Imagine that
the quark subprocess is $\bar b d \to (\bar c u)_{\bar D^0} \bar d d$, with
subsequent fragmentation of $\bar d d$ into $\bar p p$ through the creation
of two additional $u \bar u$ pairs.  The fragmentation rate for $\bar d d$ into
$\bar p p$ may differ from that into $\bar n n$ and other $B_8 \bar B_8$ pairs.
Moreover, the fact that the $d$ is a spectator quark while $\bar d$ was
produced in the weak decay can lead to kinematic asymmetries.  The Dalitz
plot need no longer be uniform along the low-mass $p \bar p$ band.  Since
both the $\bar D^0$ and $\bar d$ are produced in the weak decay, they are
correlated, leading one to expect the inequality $\langle M(\bar D^0 \bar p)
\rangle < \langle M(\bar D^0 p) \rangle$ between the average effective masses
of pairs, and $\langle p^*_{p} \rangle < \langle p^*_{\bar p} \rangle$ between
average c.m.\ momenta.

As has been pointed out elsewhere (see in particular Fig.\ 2 of the last of
Refs.\ \cite{CY}), there are other subprocesses contributing to $B^0 \to \bar
D^0 p \bar p$.  One involves the exchange process $\bar b d \to \bar c u$,
followed by the fragmentation of $\bar c u$ to $\bar D^0 p \bar p$.  Such
processes are expected to be suppressed in other $B$ decays (see, e.g.,
\cite{CR}) and there is no reason to expect them to play a major role here.

Another color-suppressed mechanism involves the quark subprocess $\bar b \to
\bar c u \bar d$ in which the $u$ is incorporated into a baryon while the $\bar
d$ is incorporated into an antibaryon.  This process is not expected to
lead to a low-mass baryon-antibaryon enhancement.  Its relative
importance is hard to estimate without a detailed flavor-symmetry analysis.

Similar arguments apply to the decay $B^0 \to \bar \Lambda p \pi^-$.  Since
strangeness-changing charmless $B$ decays appear to be dominated by the
$\bar b \to \bar s$ penguin amplitude, we assume that to be the case here,
so the expected quark subprocess is $\bar b d \to \bar s d$ followed by
fragmentation of $\bar s d$ into $\bar \Lambda p \pi^-$.  A graph for this
process, reading from top to bottom, involves the following subprocesses.

The $\bar s$ antiquark is ``dressed'' by a $\bar u \bar d$ anti-diquark to form
a $\bar \Lambda$.  The anti-diquark is produced with a $ud$ diquark which is
dressed by a $u$ quark to form a proton.  The $u$ quark is produced with a
$\bar u$ antiquark which forms a $\pi^-$ with the spectator $d$ quark.

In this process the $\bar \Lambda$ and the proton are neighbors in the
fragmentation chain.  One thus expects their effective mass to be low, as is
seen.  Since the $p$ and $\pi^-$ are also neighbors, one expects their
average effective mass $\langle M(p \pi^-) \rangle$ to be less than the average
effective mass $\langle M(\bar \Lambda \pi) \rangle$.

The observed branching ratio \cite{Lppi}
\beq \label{eqn:Lppi}
\b(B^0 \to \bar \Lambda p \pi^-) = (3.97^{+1.00}_{-0.80} \pm 0.56) \times
10^{-6}
\eeq
is quite similar to that for $B^+ \to K^+ p \bar p$ quoted in Eq.\
(\ref{eqn:Kpp}).  Thus, one might expect at least some contribution to this
last process from fragmentation.  Here the quark subprocess is expected to
be $\bar b u \to \bar s u$, followed by $\bar s u$ fragmentation into the
final state.  Reading again from top to bottom in the diagram, the $\bar s$
is dressed with a $u$ to form a $K^+$.  The $u$ is produced in a pair with
a $\bar u$.  The $\bar u$ is dressed with $\bar u \bar d$ to form a $\bar p$.
The $\bar u \bar d$ is produced with a $u d$, which combine with the spectator
$u$ to form a proton.

In the fragmentation picture for $B^+ \to K^+ p \bar p$, the Dalitz plot need
not be uniform along the low-$p \bar p$-mass band.  One expects the fact that
the $K^+$ and $\bar p$ are neighbors along the fragmentation chain to result
in $\langle M(K^+ \bar p) \rangle < \langle M(K^+ p) \rangle$ and 
$\langle p^*_p \rangle < \langle p^*_{\bar p} \rangle$.  The $\bar p p$ system
in this case has been formed by fragmentation of a $\bar u u$ pair, which is
not a flavor singlet, so there are no simple relations for production of
other baryon-antibaryon pairs.

A further example which may shed light on the fragmentation process is the
decay $B^+ \to \pi^+ \bar \Lambda_c p$ \cite{CL3b,Be3b}.  This process has a
color-suppressed
contribution which can be visualized as involving the intermediate state
$\bar \Lambda_c \Delta^{++}$ (treated recently in \cite{LR}), but more
importantly a color-favored contribution involving the subprocess $\bar b
\to \pi^+ \bar c$.  The $\bar c$ and the spectator $u$ quark then fragment
into a $\bar \Lambda_c p$ final state.  Simple kinematic arguments then favor
low $M(\bar \Lambda_c p)$, as is apparently observed \cite{SOpc}.

A similar discussion applies to the decay $B^0 \to \pi^+ \pi^- \bar \Lambda_c
p$ \cite{CL3b,Be3b}. Here the $\bar c$ produced in the color-favored subprocess
$\bar b \to \pi^+ \bar c$ combines with a spectator $d$ to produce $\pi^- \bar
\Lambda_c p$.  This system should have a low effective mass, as should its
$\bar \Sigma_c^{--} p$ component.  Another mechanism for the decay $B^0 \to
\pi^+ \pi^- \bar \Lambda_c p$ is $B^0 \to \Sigma_c^{--} \Delta^{++}$, which
proceeds only via $W$ exchange \cite{LR} and thus is expected to be highly
suppressed.

Some baryon-production processes in $B$ decays, such as $B^0 \to D^{*-} p
\bar p \pi^-$ and $B^0 \to D^{*-} p \bar n$ \cite{CLDNN}, occur with much
larger branching ratios [${\cal O}(10^{-3})$] than penguin-mediated processes
such as (\ref{eqn:Kpp}) or (\ref{eqn:Lppi}).  In these, it appears that the
charged weak current is fragmenting into a nucleon-antinucleon system
(plus possible additional pions) \cite{CHT,CY}.  Nucleon form factors then
favor low effective masses for these subsystems.

If all of the above processes are shown to be compatible with a fragmentation
process, what does one learn?  First of all, one would then have established
the phenomenological observation that fragmentation into baryon-antibaryon
pairs leads to low effective masses for those pairs.  This feature should be
taken into account in any simulation which seeks to describe baryon production.
Second, one would have established another feature of low-energy quantum
chromodynamics for which any non-perturbative approach (such as lattice gauge
theory) is obliged to provide an explanation.
\bigskip

\centerline{\bf IV.  EXOTIC RESONANCES IN $B$ DECAYS}
\bigskip

The fact that some $B$ decays lead to low-mass baryon-antibaryon enhancements
encourages the re-opening of an old question which has never been
satisfactorily addressed:  If such enchancements {\it do} exist, are they
limited to the ordinary quantum numbers of the $q \bar q$ system?  Some
arguments based on duality \cite{JRbb,HHdd,JRdd} or the systematics of
resonance formation \cite{JRcrf} suggest instead that baryon-antibaryon
enhancements are possible in all systems with the quantum numbers of
{\it two} quarks and {\it two} antiquarks.  If such resonances exist, why
aren't they seen in ordinary meson-meson channels?  A consistent set of
selection rules was proposed \cite{FWR} to forbid such mesonic couplings.
$B$ decays offer a new opportunity to test such rules.

Let us consider the decay of a $B^+$ at the quark level:  $\bar b u \to
\bar c u \bar d u$.  The final state is ``exotic'' in the sense that it
does not share flavor quantum numbers with any quark-antiquark state.
Now let the charmed antiquark $\bar c$ fragment into a $D^-$ by dressing
itself with a $d$ quark.  This is produced in a pair with a $\bar d$, so
that in addition to the $D^-$ we have a meson with the quark content $\bar d
\bar d u u$.  This is an exotic meson.

We thus suggest that in the decay $B^+ \to D^- X^{++}$ the missing mass of
$X^{++}$ be studied.  If the selection rules of Ref.\ \cite{FWR} are valid,
any resonances in the $X^{++}$ channel should decay to baryon-antibaryon
pairs.  Such pairs might be $p \bar \Delta^+$, $\Delta^{++} \bar n$, or
$\Delta^{++} \bar \Delta^0$.  The last final state has the property that
$p \bar p \pi^+ \pi^+$ is one of its decay products; the others involve
antineutrons and thus might be tricky to observe.

If the $\bar c$ quark instead fragments to a $D_s^-$ by dressing itself
with a $s$ quark, the remaining meson has the quark content $\bar s \bar d
u u$.  Thus in $B^+ \to D_s^- X^{++}$ if the missing-mass of $X^{++}$ displays
peaks, one should see whether such resonances decay to baryon-antibaryon pairs
such as $\bar \Lambda p \pi^+$.

The selection rules of Ref.\ \cite{FWR} also imply that the $\bar c \bar d
u u$ system produced by a $B^+$ decay can fragment into an exotic antibaryon
(composed of four antiquarks and a quark) and a baryon.  All one needs is the
production of two extra pairs, $\bar q_1 \bar q_2 q_1 q_2$, where neither $q_1$
nor $q_2$ is a $u$ quark.  Then the exotic antibaryon is $\bar c \bar d \bar
q_1 \bar q_2 u$, while the baryon is $u q_1 q_2$.  If $q_1 = q_2 = d$, the
baryon is a neutron.  The system $X^+$ in $B^+ \to X^+ n$ is exotic, but the
neutron is difficult to detect.  If $q_1 = d$ and $q_2 = s$, the baryon can
be a $\Lambda$ (easier to see).  The system $X^+$ in $B^+ \to X^+ \Lambda$
again is exotic; a missing-mass plot would be interesting.  Depending on the
relative masses of exotic baryons and exotic mesons, such a state might be
forced to decay via a violation of the selection rules of Ref.\ \cite{FWR}.
\bigskip

\centerline{\bf V.  CONCLUSIONS}
\bigskip

The observation of low-mass baryon-antibaryon enhancements in $B$ decays has
opened a range of interesting possibilities.  Some of these enhancements may
be associated with coupling to flavorless states of two or more gluons, and
may be related to the enhanced branching ratios for $B \to \eta' K$ and
$B \to \eta' X$.  If they are associated with spinless versions of such
states, specific features of the Dalitz plots for three-body decays are
expected.  Other enhancements may be associated with details of the
fragmentation picture, suggesting a short-range correlation between baryons
and antibaryons in the fragmentation chain.  The possibility that exotic
mesons and baryons may be observable in the decays of charged $B$ mesons is a
further outcome of the recent experimental observations.  
\bigskip

\centerline{\bf ACKNOWLEDGMENTS}
\bigskip

I am grateful to Tom Browder, Steve Olsen, Sandip Pakvasa, Xerxes Tata, and
San Fu Tuan for discussions, and to the Physics Department of the University
of Hawaii for hospitality during part of this research.  This work was
supported in part by the United States Department
of Energy through Grant No.\ DE FG02 90ER40560.

\def \ajp#1#2#3{Am.\ J. Phys.\ {\bf#1}, #2 (#3)}
\def \apny#1#2#3{Ann.\ Phys.\ (N.Y.) {\bf#1}, #2 (#3)}
\def \app#1#2#3{Acta Phys.\ Polonica {\bf#1}, #2 (#3)}
\def \arnps#1#2#3{Ann.\ Rev.\ Nucl.\ Part.\ Sci.\ {\bf#1}, #2 (#3)}
\def \art{and references therein}
\def \cmts#1#2#3{Comments on Nucl.\ Part.\ Phys.\ {\bf#1}, #2 (#3)}
\def \cn{Collaboration}
\def \cp89{{\it CP Violation,} edited by C. Jarlskog (World Scientific,
Singapore, 1989)}
\def \efi{Enrico Fermi Institute Report No.\ }
\def \epjc#1#2#3{Eur.\ Phys.\ J. C {\bf#1}, #2 (#3)}
\def \f79{{\it Proceedings of the 1979 International Symposium on Lepton and
Photon Interactions at High Energies,} Fermilab, August 23-29, 1979, ed. by
T. B. W. Kirk and H. D. I. Abarbanel (Fermi National Accelerator Laboratory,
Batavia, IL, 1979}
\def \hb87{{\it Proceeding of the 1987 International Symposium on Lepton and
Photon Interactions at High Energies,} Hamburg, 1987, ed. by W. Bartel
and R. R\"uckl (Nucl.\ Phys.\ B, Proc.\ Suppl., vol.\ 3) (North-Holland,
Amsterdam, 1988)}
\def \ib{{\it ibid.}~}
\def \ibj#1#2#3{~{\bf#1}, #2 (#3)}
\def \ichep72{{\it Proceedings of the XVI International Conference on High
Energy Physics}, Chicago and Batavia, Illinois, Sept. 6 -- 13, 1972,
edited by J. D. Jackson, A. Roberts, and R. Donaldson (Fermilab, Batavia,
IL, 1972)}
\def \ijmpa#1#2#3{Int.\ J.\ Mod.\ Phys.\ A {\bf#1}, #2 (#3)}
\def \ite{{\it et al.}}
\def \jhep#1#2#3{JHEP {\bf#1}, #2 (#3)}
\def \jpb#1#2#3{J.\ Phys.\ B {\bf#1}, #2 (#3)}
\def \lg{{\it Proceedings of the XIXth International Symposium on
Lepton and Photon Interactions,} Stanford, California, August 9--14 1999,
edited by J. Jaros and M. Peskin (World Scientific, Singapore, 2000)}
\def \lkl87{{\it Selected Topics in Electroweak Interactions} (Proceedings of
the Second Lake Louise Institute on New Frontiers in Particle Physics, 15 --
21 February, 1987), edited by J. M. Cameron \ite~(World Scientific, Singapore,
1987)}
\def \kdvs#1#2#3{{Kong.\ Danske Vid.\ Selsk., Matt-fys.\ Medd.} {\bf #1},
No.\ #2 (#3)}
\def \ky85{{\it Proceedings of the International Symposium on Lepton and
Photon Interactions at High Energy,} Kyoto, Aug.~19-24, 1985, edited by M.
Konuma and K. Takahashi (Kyoto Univ., Kyoto, 1985)}
\def \mpla#1#2#3{Mod.\ Phys.\ Lett.\ A {\bf#1}, #2 (#3)}
\def \nat#1#2#3{Nature {\bf#1}, #2 (#3)}
\def \nc#1#2#3{Nuovo Cim.\ {\bf#1}, #2 (#3)}
\def \nima#1#2#3{Nucl.\ Instr.\ Meth. A {\bf#1}, #2 (#3)}
\def \np#1#2#3{Nucl.\ Phys.\ {\bf#1}, #2 (#3)}
\def \npbps#1#2#3{Nucl.\ Phys.\ B Proc.\ Suppl.\ {\bf#1}, #2 (#3)}
\def \os{XXX International Conference on High Energy Physics, Osaka, Japan,
July 27 -- August 2, 2000}
\def \PDG{Particle Data Group, K. Hagiwara \ite, \prd{66}{010001}{2002}}
\def \pisma#1#2#3#4{Pis'ma Zh.\ Eksp.\ Teor.\ Fiz.\ {\bf#1}, #2 (#3) [JETP
Lett.\ {\bf#1}, #4 (#3)]}
\def \pl#1#2#3{Phys.\ Lett.\ {\bf#1}, #2 (#3)}
\def \pla#1#2#3{Phys.\ Lett.\ A {\bf#1}, #2 (#3)}
\def \plb#1#2#3{Phys.\ Lett.\ B {\bf#1}, #2 (#3)}
\def \pr#1#2#3{Phys.\ Rev.\ {\bf#1}, #2 (#3)}
\def \prc#1#2#3{Phys.\ Rev.\ C {\bf#1}, #2 (#3)}
\def \prd#1#2#3{Phys.\ Rev.\ D {\bf#1}, #2 (#3)}
\def \prl#1#2#3{Phys.\ Rev.\ Lett.\ {\bf#1}, #2 (#3)}
\def \prp#1#2#3{Phys.\ Rep.\ {\bf#1}, #2 (#3)}
\def \ptp#1#2#3{Prog.\ Theor.\ Phys.\ {\bf#1}, #2 (#3)}
\def \rmp#1#2#3{Rev.\ Mod.\ Phys.\ {\bf#1}, #2 (#3)}
\def \rp#1{~~~~~\ldots\ldots{\rm rp~}{#1}~~~~~}
\def \rpp#1#2#3{Rep.\ Prog.\ Phys.\ {\bf#1}, #2 (#3)}
\def \sing{{\it Proceedings of the 25th International Conference on High Energy
Physics, Singapore, Aug. 2--8, 1990}, edited by. K. K. Phua and Y. Yamaguchi
(Southeast Asia Physics Association, 1991)}
\def \slc87{{\it Proceedings of the Salt Lake City Meeting} (Division of
Particles and Fields, American Physical Society, Salt Lake City, Utah, 1987),
ed. by C. DeTar and J. S. Ball (World Scientific, Singapore, 1987)}
\def \slac89{{\it Proceedings of the XIVth International Symposium on
Lepton and Photon Interactions,} Stanford, California, 1989, edited by M.
Riordan (World Scientific, Singapore, 1990)}
\def \smass82{{\it Proceedings of the 1982 DPF Summer Study on Elementary
Particle Physics and Future Facilities}, Snowmass, Colorado, edited by R.
Donaldson, R. Gustafson, and F. Paige (World Scientific, Singapore, 1982)}
\def \smass90{{\it Research Directions for the Decade} (Proceedings of the
1990 Summer Study on High Energy Physics, June 25--July 13, Snowmass, Colorado),
edited by E. L. Berger (World Scientific, Singapore, 1992)}
\def \tasi{{\it Testing the Standard Model} (Proceedings of the 1990
Theoretical Advanced Study Institute in Elementary Particle Physics, Boulder,
Colorado, 3--27 June, 1990), edited by M. Cveti\v{c} and P. Langacker
(World Scientific, Singapore, 1991)}
\def \yaf#1#2#3#4{Yad.\ Fiz.\ {\bf#1}, #2 (#3) [Sov.\ J.\ Nucl.\ Phys.\
{\bf #1}, #4 (#3)]}
\def \zhetf#1#2#3#4#5#6{Zh.\ Eksp.\ Teor.\ Fiz.\ {\bf #1}, #2 (#3) [Sov.\
Phys.\ - JETP {\bf #4}, #5 (#6)]}
\def \zpc#1#2#3{Zeit.\ Phys.\ C {\bf#1}, #2 (#3)}
\def \zpd#1#2#3{Zeit.\ Phys.\ D {\bf#1}, #2 (#3)}

\end{document}